# Observation of Anomalous Thermal Hall Effect in a Kagome Superconductor

Hiroki Yoshida[1], Hikaru Takeda[1], Jian Yan[1,2], Yui Kanemori[1], Brenden R. Ortiz[3], Yuzki M. Oey[4], Stephen D. Wilson[4], Marcin Konczykowski[5], Kota Ishihara[6], Takasada Shibauchi[6] and Minoru Yamashita[1*]

1 The Institute for Solid State Physics, The University of Tokyo, Kashiwa, Chiba 277–8581, Japan
2 Institute for Advanced Study, Chengdu University, Chengdu 610106, China
3 Materials Science and Technology Division, Oak Ridge National Laboratory, Oak Ridge, TN, USA
4 Materials Department, University of California Santa Barbara, Santa Barbara, CA 93106, USA
5 Laboratoire des Solides Irradiés, CEA/DRF/IRAMIS, Ecole Polytechnique, CNRS, Institut Polytechnique de Paris, F-91128 Palaiseau, France
6 Department of Advanced Materials Science, The University of Tokyo, Kashiwa, Chiba 277-8561, Japan
* Corresponding author. E-mail: my@issp.u-tokyo.ac.jp

**Abstract**

Broken time-reversal symmetry (TRS) in superconductors can induce not only spontaneous magnetization by the finite angular momentum of Cooper pairs, but also anomalous thermal Hall effects (ATHEs), whose detection has been extremely challenging. Here we report the successful observation of an ATHE developing below the superconducting transition temperature at zero magnetic field in the kagome-lattice superconductor $CsV_3Sb_5$. This finding is verified by the absence of a signal in a conventional type-II superconductor using the same setup and by ruling out the trapped-vortex effects through micro-Hall array measurements. Remarkably, both the temperature dependence and the magnitude of the observed anomalous thermal Hall conductivity are quite different from those expected for the quantized thermal edge current of an intrinsic ATHE, but consistent with extrinsic impurity-induced ATHEs in chiral superconductivity. Our study of ATHE offers an alternative approach to probe TRS breaking in the superconducting states.

Introduction:

Spontaneous time-reversal symmetry (TRS) breaking is one of the interesting cooperative phenomena brought by a phase transition in solids, which allows both a spontaneous magnetic moment as observed in a ferromagnet and a chiral edge flow to be realized. For example, in a magnetic topological insulator, the dissipationless quantized edge current is realized even in zero field owing to the anomalous quantum Hall effect by the broken TRS(*1, 2*). A chiral ground state of localized spins breaking the TRS may also exhibit a spontaneous Hall effect, as reported in the metallic spin ice compound(*3*).

Of particular interest in a chiral flow by the broken TRS is the topologically-protected edge current in a chiral superconductor(*4–6*) in which spontaneous magnetic moment appears by forming Cooper pairs with a finite orbital angular momentum. In contrast, conventional superconductivity is incompatible with collective magnetism and broken TRS. Although a charge current is not conserved in a superconductor, a thermal current brought by Bogoliubov quasiparticles is conserved, giving rise to a quantization of the thermal Hall conductivity in a chiral superconductor by the Chern number characterizing the chiral superconducting state(*7–9*). In addition to such an intrinsic anomalous thermal Hall effect (ATHE), extrinsic ATHEs by impurity scatterings are also suggested in chiral superconductors(*10–17*). Although numerous candidate materials have been reported so far to realize chiral superconductivity, mainly by the observation of the spontaneous magnetization by the polar Kerr or the μSR measurements(*6*), these ATHEs have yet to be observed.

Here, we report the successful observation of anomalous thermal Hall conductivity ($\kappa_{xy}^{\mathrm{ATHE}}$) emerging at zero external field below the superconducting transition temperature in the kagome superconductor $CsV_3Sb_5$(*18–20*). The anomalous thermal Hall conductivity is determined by measuring the transverse temperature difference ($\Delta T_y$) caused by a thermal current ($Q$) at zero external field, after cooling the sample under a finite magnetic field at the superconducting transition to polarize the domains of the chiral superconductor. Reversing this "training" field ($B_{\mathrm{tr}}$) allows us to estimate $\kappa_{xy}^{\mathrm{ATHE}}$ by antisymmetrizing the transverse thermal-Hall resistance to remove the mixed longitudinal component. We verify our experimental setup by confirming the null result for the conventional type-II superconductor 2$H$–NbS$_2$. We further exclude the effects of trapped fluxes in the sample by measuring the training-field dependence of the trapped field by micro-Hall array measurements. Our results demonstrate that the magnitude of $\kappa_{xy}^{\text{ATHE}}/T$ observed in CsV$_3$Sb$_5$ exceeds the theoretical value expected for an intrinsic ATHE of a chiral superconductor by more than one order of magnitude, as well as exhibits a different temperature dependence of $\kappa_{xy}^{\text{ATHE}}/T$ from that of an intrinsic ATHE(*7, 9*). On the other hand, both the magnitude and the temperature dependence of $\kappa_{xy}^{\text{ATHE}}/T$ are consistent with an extrinsic impurity-induced ATHE that predicts a temperature dependence of $\kappa_{xy}^{\text{ATHE}}/T$ with a peak at a fraction of the superconducting transition temperature without the residual of $\kappa_{xy}^{\text{ATHE}}/T$ depending on the impurities(*13–17*), suggesting observation of extrinsic ATHE in a chiral superconducting state in CsV$_3$Sb$_5$. The method we use to observe the ATHE is applicable to various superconductors, which will bring substantial advances in the research on chiral superconductivity.

Results:

The kagomé superconductor CsV$_3$Sb$_5$, which consists of the covalent V-Sb sheets sandwiched by the honeycomb lattice of Cs intercalant(*18–20*) (Fig. 1A), shows the superconducting transition at $T_c = 3.5$ K inside the charge density wave (CDW) phase formed below $T_{\text{CDW}} = 94$ K. The possible breaking of the TRS has been discussed in the superconducting phase(*21, 22*), the CDW phase(*23–26*) or even above $T_{\text{CDW}}$(*27*), although some negative results have also been reported(*28*). The kagome network of the V atoms results in the electronic band structure with the Van Hove singularities around the M point and the Dirac crossing at the K point. Although chiral $d$-wave superconductivity is theoretically suggested to be realized in such a kagome metal(*29*), the experimental investigations about the superconducting gap structure of CsV$_3$Sb$_5$ are converging to indicate an anisotropic $s$-wave, multi-gap structure without a node(*30–36*).

We investigate the electrical and thermal transport properties of single crystals of CsV$_3$Sb$_5$ by using the setup shown in Fig. 1B. The temperature dependence of the resistivity ($\rho$) at zero field shows a clear superconducting transition with the zero resistivity observed below $T_{c0} = 3.0$ K (Fig. 1C). This superconducting transition is also observed in the temperature dependence of the heat capacity divided by the temperature ($C/T$) as the transition peak with the onset around 2.6 K (Fig. 1D). This superconducting transition temperature is comparable or even higher than that reported in previous works(*19, 35, 36*), showing no notable suppression of the spin-singlet superconductivity by magnetic impurities. Moreover, the temperature dependence of the resistivity in the normal state (Fig. 1C) shows no sign of a Kondo peak associated with magnetic impurities. These results ensure the high quality of the sample with negligible magnetic impurities. The superconductivity can be suppressed by applying the magnetic field above the irreversibility field ($B_{\mathrm{irr}}$) around 0.5 T (Fig. 2A) along the $c$ axis. Figure 1E shows the temperature dependence of the longitudinal thermal conductivity divided by the temperature ($\kappa_{xx}/T$) at zero field and in the normal state at 1.5 T, well reproducing previous work(*37*). The dotted line in Fig. 1E shows the electronic contribution in $\kappa_{xx}/T$ estimated from the resistivity data in the normal state at 1.5 T (Fig. 1C) by the Wiedemann-Franz law. The temperature dependence of $\kappa_{xx}/T$ at the same field asymptotically approaches the Wiedemann-Franz value at lower temperatures, showing a good thermalization between the electrons and phonons. The excess $\kappa_{xx}/T$ beyond the Wiedemann-Franz value indicates the phonon contribution to the thermal conduction of this compound.

Although an anomalous thermal Hall effect is a phenomenon emerging at zero field, a zero-field cooling of the sample would result in multiple random domains of the orbital angular momentum of the Cooper pairs. Therefore, to polarize the domains, the sample was first cooled under a finite magnetic field, denoted as a "training field" $B_{\mathrm{tr}}$, across the superconducting transition. At the lowest temperature, the training magnetic field was turned off by monitoring with the Hall sensor placed near the sample (the blue arrow in Fig. 2A, see also Methods and section S1 in the Supplementary Materials (SM) for determining the superconducting phase diagram). Then, in the zero applied-field condition, the temperature dependence of the thermal Hall resistivity $\lambda_{xy}(T, +B_{\mathrm{tr}}) = t \cdot$ $\Delta T_y/Q$, where $t$ is the thickness of the sample, is measured with increasing temperature above $T_c$ (the green arrow in Fig. 2A). A representative data of $\lambda_{xy}$ recorded for $B_{tr} = 10$ mT is shown in Fig. 2B. As shown in Fig. 2B, the temperature dependence of $\lambda_{xy}$ is almost identical with that of the longitudinal thermal resistivity ($\lambda_{xx}$), showing the parasitic contribution of $\lambda_{xx}$ mixed in $\lambda_{xy}$ due to the imperfect placement of the transverse thermometers ($T_{L1}$ and $T_{L2}$) against perpendicular to the heat current (see section S2 in SM for more details about the mixing). To remove this mixture, the same procedure was repeated with the reversed $B_{tr}$ (the cyan arrow in Fig. 2A) to antisymmetrize the thermal Hall resistivity as $\lambda_{xy}^{\text{asym}} = [\lambda_{xy}(T, +B_{tr}) - \lambda_{xy}(T, -B_{tr})]/2$, as done for the Hall-voltage measurements. We stress that, however, in contrast to the ordinal Hall measurements done with reversing the magnetic fields, only $B_{tr}$ in the cooling process across the superconducting transition was reversed in our anomalous thermal Hall measurements; all the $\lambda_{xy}$ measurements were performed under the zero external magnetic field (or in the small remnant field as discussed below). As shown in Fig. 2C, this antisymmetrization allows us to resolve the finite $\lambda_{xy}^{\text{asym}}$ that develops below $\sim T_{c0}/2$, demonstrating the emergence of the spontaneous thermal Hall effect in the superconducting phase. We note that $\lambda_{xx}$ also contains a field-antisymmetric component that cannot be explained solely by a mixing of $\lambda_{xy}$, which however does not affect the estimation of $\lambda_{xy}^{\text{asym}}$. See section S2 in SM.

The temperature dependence of $\kappa_{xy}^{\text{ATHE}} = \lambda_{xy}^{\text{asym}} \cdot \kappa_{xx}^2$ observed after cooling under different $B_{tr}$ is summarized in Fig. 2D. As shown in Fig. 2D, $\kappa_{xy}^{\text{ATHE}}/T$ starts to increase below about 2 K, which is followed by the peak around 1 K and the rapid decrease to zero in the lower temperatures. We find that the temperature dependence of $\kappa_{xy}^{\text{ATHE}}/T$ does not depend on $B_{tr}$ within the noise level of our ATHE measurements in the whole training field range we measured. We also investigate the stability of $\kappa_{xy}^{\text{ATHE}}$ by intentionally setting the finite remnant field ($B_{\text{rem}}$) after the cooling at $B_{tr}$ (the orange arrow in Fig. 3A), of which the data was antisymmetrized by measuring $\lambda_{xy}$ at $-B_{\text{rem}}$ after the cooling at $-B_{tr}$. As shown in Fig. 3B, $\kappa_{xy}^{\text{ATHE}}$ is not affected for $|B_{\text{rem}}| < 1$ mT, ensuring that $\kappa_{xy}^{\text{ATHE}}$ is not affected by a small residual field ($<1$ mT) at the zero-field $\lambda_{xy}$ measurements. On the other hand, $\kappa_{xy}^{\text{ATHE}}$ shows a sign reversal for $B_{\text{rem}} <$ $-2$ mT, implying the reversal of the polarity of the chiral domains (see also section S6 in SM). It should be noted that a larger magnetic field is required to flip the polarization of the superconducting state at 0.5 K compared to that required for the training field. This is because the training field is applied during the superconducting transition at which the energy required to flip the polarization should be zero because of the zero condensation energy at $T_c$. On the other hand, the remnant field is applied at 0.5 K ($\sim T_c/6$) at which the condensation energy is almost fully developed(*30*, *35*, *36*). We also note that a similar temperature dependence of $\kappa_{xy}^{\text{ATHE}}$ is observed in a different single crystal of CsV$_3$Sb$_5$ (Fig. S4 in SM), confirming the good reproducibility of $\kappa_{xy}^{\text{ATHE}}$.

We scrutinize the observation of $\kappa_{xy}^{\text{ATHE}}$ in several ways. First, we have repeated the same measurements by using the same setup on the conventional type-II superconductor 2*H*–NbS$_2$(*38*, *39*). We find that $\kappa_{xy}^{\text{ATHE}}$ in 2*H*–NbS$_2$ is negligibly small compared to that in CsV$_3$Sb$_5$ as shown in Fig. 4. The same result can be obtained by comparing the thermal Hall angle data (see section S4 for more results obtained in 2*H*–NbS$_2$) to take into account the difference in the magnitude of $\kappa_{xx}$. This null result in 2*H*–NbS$_2$ ensures that $\kappa_{xy}^{\text{ATHE}}$ observed in CsV$_3$Sb$_5$ is inherent to the superconducting state of this material, not an artifact of our experimental setup.

Second, we have investigated the effect of the trapped fluxes owing to the training field cooling into the superconducting phase. The amount of the trapped fluxes was checked by using the micro-Hall array(*40*) placed directly under the sample (Figs. 5A–C). After repeating exactly the same procedure with $B_{\text{tr}}$ as the $\kappa_{xy}^{\text{ATHE}}$ measurements, the temperature dependence of the trapped fluxes was measured at zero field by using the micro-Hall sensor located near the center of the sample, at which the trapped field flux density is expected to be maximum. As shown in Fig. 5D, the trapped field at the lowest temperature is increased as $B_{\text{tr}}$ increases up to about 3 mT, which saturates for $B_{\text{tr}} > 3$ mT due to the weak vortex pinning(*41*). We confirm that this trapped field is not affected by a thermal gradient applied as when the thermal-transport measurements were done (see section S5 in SM). We further find that the trapped field becomes slightly asymmetric with respect to the field direction when $B_{\text{tr}} > 1$ mT (Fig. S6 in SM), possibly due to an asymmetric critical current by the broken TRS in the superconducting state.

Despite this clear $B_{\text{tr}}$ dependence of the trapped field, the temperature dependence of $\kappa_{xy}^{\text{ATHE}}$ in CsV$_3$Sb$_5$ stays similar below 3 mT (Fig. 5E), verifying that an effect by the trapped fluxes on $\kappa_{xy}^{\text{ATHE}}$ is smaller than the noise level of our ATHE measurements. Therefore, the trapped fluxes can be ruled out as the cause of the observed $\kappa_{xy}^{\text{ATHE}}$. This irrelevance may suggest that the heat carriers under the influence of the trapped vortices are localized inside the fluxes, and thus do not contribute to the overall thermal Hall effect unless the trapped fluxes overlap each other, as is known for the field dependence of the longitudinal thermal conduction in fully-gapped superconductors(*42*). Indeed, the inter-vortex distance of about 1.5 μm, which is estimated for the trapped field of 3 mT by assuming a triangular lattice of the trapped vortices, is much longer than the penetration depth of 200–400 nm of CsV$_3$Sb$_5$(*30, 33*).

## Discussion:

We first examine the possibility that the observed ATHE is caused by an intrinsic thermal Hall effect given by the Chern number(*7–9*) characterizing the chiral superconductivity suggested in this compound(*21, 22*). Similar to a topological insulator exhibiting a quantized Hall conductivity given by $\sigma_{xy} = C\, e^2/h$, a chiral superconductor shows the spontaneous thermal Hall conductivity quantized by the Chern number $C$ which characterizes the chiral superconducting state as(*7–9*),

$$\frac{\kappa_{xy}^{\text{spnt}}}{T} = \frac{C}{2} K_0, \quad (1)$$

where $K_0 = \frac{\pi}{6}\frac{k_B^2}{\hbar}$ is the quantum thermal conductance and the factor of 1/2 represents the Majorana edge state of the thermal flow(*4, 7*). Although the Chern number of the superconducting state in CsV$_3$Sb$_5$ is unknown, this theoretical estimate for $C = 1$ gives $\kappa_{xy}^{\text{spnt}}/T = \frac{\pi}{12}\frac{k_B^2}{\hbar}\frac{1}{d} \sim 0.0005$ W K$^{-2}$ m$^{-1}$, where $d = 0.92$ nm is the interlayer distance(*19*). Therefore, this theoretical prediction is more than one order of magnitude smaller than the peak magnitude of $\kappa_{xy}^{\text{ATHE}}/T \sim 0.02$ W K$^{-2}$ m$^{-1}$ observed in CsV$_3$Sb$_5$.

Moreover, the temperature dependence of $\kappa_{xy}^{\text{ATHE}}/T$ in CsV$_3$Sb$_5$ shows a peak around 1 K with zero residual in the zero-temperature extrapolation, which is quite different from the theoretical calculation of the Bogoliubov quasiparticles in a chiral superconductor showing a monotonic increase of $\kappa_{xy}^{\text{spnt}}/T$ upon lowering the temperature with the finite residual given by Eq. (1)(*4*, *7*, *9*). The larger $\kappa_{xy}^{\text{ATHE}}/T$ at the peak temperature around 1 K might be explained as an enhancement effect by a coupling between the Bogoliubov quasiparticles and the phonons as discussed in a Kitaev magnet where an enhancement of $\kappa_{xy}$ is shown to appear by a coupling between the Majorana fermions and the phonons(*43*, *44*). However, the absence of the residual $\kappa_{xy}^{\text{ATHE}}/T$ in the zero temperature limit is inconsistent with the intrinsic mechanism, suggesting that $\kappa_{xy}^{\text{ATHE}}$ observed in CsV$_3$Sb$_5$ is due to a different origin.

Next, we examine an extrinsic impurity-induced ATHE for $\kappa_{xy}^{\text{ATHE}}/T$ observed in CsV$_3$Sb$_5$. Like anomalous Hall effects in ferromagnetic metals(*45*), not only an intrinsic but also impurity-induced extrinsic mechanisms(*10–17*) have been suggested as ATHEs in a chiral superconductor. In an extrinsic ATHE, impurity-scattering effects on the chiral condensates by non-magnetic impurities give rise to $\kappa_{xy}^{\text{ATHE}}$ that depends on both the characteristics of the chiral order parameter and the details of the impurity scatterings. Remarkably, depending on the details of the impurities, the magnitude of impurity-induced ATHEs have been shown to become orders of magnitude larger than the intrinsic quantized edge contribution given by Eq. (1) and to exhibit a similar temperature dependence to the experimental result with a peak at a fraction of $T_c$. Indeed, it has been shown that $\kappa_{xy}^{\text{ATHE}}$ of chiral spin-singlet state belonging to $E_{1g}$ and $E_{2g}$ representations easily become the size of the observed $\kappa_{xy}^{\text{ATHE}}$ for small impurity density(*13*, *14*, *16*). For example, the calculation of $\kappa_{xy}^{\text{ATHE}}$ for a chiral *d*-wave state shown in Ref. (*14*) shows that the peak of the observed thermal Hall signal, $\kappa_{xy}^{\text{ATHE}} \sim 0.005\kappa_{xx}(T_c)$, is given by the impurity size of $\sim 0.5/k_F$, where $k_F$ is the Fermi wave number. Impurities of such atomic size are likely to be present in this compound. We note that the mean free path used in the calculation $\ell_{\text{mf}}/\xi_0 = 7.5$, where $\xi_0$ is the coherence length, is in the same order with that estimated in our sample of $\ell_{\text{mf}}/\xi_0 \sim 20$–$40$ estimated by the residual resistivity (see section S7 in SM). Therefore, it is likely that the observed $\kappa_{xy}^{\mathrm{ATHE}}$ in CsV$_3$Sb$_5$ is caused by an extrinsic ATHE realized in a chiral superconducting state.

In contrast to the intrinsic chiral edge current that is characterized surely by the Chern number of the chiral superconductivity, it is not straight forward to pin down the chiral superconducting state from the extrinsic ATHE measurements because extrinsic ATHEs strongly depend on the details of the impurity scattering (scattering strength, impurity size, etc.). According to the theoretical calculations(*13*, *14*, *16*, *17*), the intrinsic component becomes dominant over the extrinsic one only at very low temperatures ($T \ll T_\mathrm{c}$) in a clean chiral superconductor without an in-gap bound state at the Fermi energy. Therefore, to search for the intrinsic ATHE in the future, it will be necessary to achieve $\kappa_{xy}^{\mathrm{ATHE}}$ measurements down to lower temperatures in samples with fewer impurities.

It should be noted that the study of the impurity effects on the superconducting gap of CsV$_3$Sb$_5$ precludes a sign-changing order parameter including chiral states(*32*, *35*, *36*). On the other hand, the recent theoretical study points out that a chiral spin-singlet state in CsV$_3$Sb$_5$ can be robust against point-like impurities on the V sites due the particular sub-lattice character of the kagome structure(*46*). Meanwhile, it has been shown that extrinsic ATHEs are not caused by a point-like impurity, but require finite-size impurities for chiral superconductors with the winding number greater than or equal to two, including a chiral *d*-wave state(*13*, *14*). Therefore, the impurity providing the observed $\kappa_{xy}^{\mathrm{ATHE}}$ could be different from the point-like impurities discussed in terms of the phase-sensitive measurements. Further theoretical studies to estimate effects of the finite-size impurities on both the suppression of $T_\mathrm{c}$ and the magnitude of $\kappa_{xy}^{\mathrm{ATHE}}$ based on the electronic state of the kagome structure are required to clarify the superconducting order parameter of this compound from our $\kappa_{xy}^{\mathrm{ATHE}}$ data.

The possible origin of the observed ATHE is not limited to the Bogoliubov quasiparticles in a chiral superconductor but includes any quasiparticles in TRS-breaking electronic orders. For instance, the charge-loop current orders with broken TRS have been proposed in CsV$_3$Sb$_5$, where the current direction can be controlled by applying the magnetic field(*26, 27*). Given the interplay between the charge-loop current order and bond-order fluctuations suggested in this compound(*47, 48*), the charge-loop current, whose domains are aligned by the training field, may couple with the phonon heat conduction through the bond fluctuations. Since the opening of superconducting gap is known to extend the quasiparticle lifetime below $T_c$, the spontaneous thermal Hall effect by the charge-loop current might be enlarged in the superconducting phase due to the extension of the mean free path.

Our measurements of the anomalous thermal Hall effect at zero external field provide an alternative approach to investigate TRS breaking in the superconducting state, which can be applied to a wide range of candidate materials of chiral superconductivity(*6*). The observed anomalous thermal Hall effect in CsV$_3$Sb$_5$ is likely caused by an extrinsic impurity-scattering effect on the chiral condensates, facilitating further studies of the intrinsic and extrinsic anomalous thermal Hall effect in chiral superconductors.

Materials and Methods:

The single crystals of CsV$_3$Sb$_5$ were synthesized using the self-flux method as described in the previous works(*18, 19*). Two single crystals (#1 and #2) were used for the electrical- and the thermal-transport measurements. The results obtained in sample #1 (#2) are shown in the main text (Fig. S4 in SM). The single crystal of 2*H*–NbS$_2$ was purchased from HQ Graphene. The sample dimensions of these samples are listed in Table S1 in SM.

The electrical- and thermal-transport measurements were performed by using the setup shown in Fig. 1B installed in a dilution refrigerator. Three calibrated RuO$_2$ thermometers ($T_{\text{High}}$, $T_{\text{L1}}$ and $T_{\text{L2}}$) and one heater were attached to the sample by using the gold wires spot-welded to the sample. These thermal contacts were also used as the electrical contact to measure the resistivity. The longitudinal thermal resistivity $\lambda_{xx}$ and the thermal Hall resistivity $\lambda_{xy}$ were determined by measuring $\Delta T_x = T_{\text{High}} - T_{\text{L1}}$ and $\Delta T_y = T_{\text{L1}} - T_{\text{L2}}$ as a function of the applied heater ($Q$) as $\lambda_{xx} = \frac{\Delta T_x}{l}\frac{wt}{Q}$ and $\lambda_{xy} = \frac{\Delta T_y}{w}\frac{wt}{Q}$ by applying a thermal gradient about $\Delta T_x/T = 1.8$–3.0%, where $l$ is the distance between the thermal contacts for $T_{\text{High}}$ and $T_{\text{L1}}$, $w$ is the width, and $t$ is the thickness of the sample, respectively (Fig. 1B). The sample temperature $T$ is determined by $T = (T_{\text{High}} + T_{\text{L1}})/2$.

The magnetic field at the sample position was monitored by the chip Hall sensor placed near the sample. Since a residual magnetic field up to about 3 mT remains in the superconducting magnet even if the output of the magnet power supply is set to 0 (see section S6 in SM for more details), the zero field during the anomalous thermal Hall measurements was achieved by adjusting the magnet power supply output so that the Hall sensor reading becomes 0. The residual magnetic field limited by the sensitivity of the Hall sensor is less than 50 µT, which is similar to the geomagnetic field in our building.

The micro-Hall array was fabricated in a GaAs/AlGaAs heterostructure, in which ten 5 µm×5 µm sized sensors are fabricated with 20 µm interval. The trapped magnetic field data (Fig. 5D) was taken using a sensor located approximately at the center of the sample. The effect of a thermal current on trapped magnetic field was investigated by attaching a heater and thermometers on the sample placed on the micro-Hall sensor (section S5 in SM).

1276 (2011).

Acknowledgements:

We thank Y. Kasahara, H. Kontani, Y. Matsuda, T. Matsushita and R. Tazai for useful discussions. We acknowledge Y. Hashimoto in Q-NanoLabo in ISSP, The University of Tokyo for his technical assistance. We thank V. Mosser for his help in design and fabrication of micro-Hall array sensors.

Funding:

This work was supported by Grants-in-Aid for Scientific Research (KAKENHI) (numbers JP22KF0111, JP23K25813, JP22H00105, and JP23H00089) and Grant-in-Aid for Transformative Research Areas (A) "Correlation Design Science" (KAKENHI Grant No. JP25H01248) from JSPS. J.Y. was supported by Grant-in-Aid for JSPS Fellows. S.D.W., B.R.O., and Y.M.O. gratefully acknowledge support via the UC Santa Barbara NSF Quantum Foundry funded via the Q-AMASE-i program under award DMR-1906325. Work by B.R.O. was supported by the U.S. Department of Energy (DOE), Office of Science, Basic Energy Sciences (BES), Materials Sciences and Engineering Division.

Author contributions:

H.Y. and M.Y. conceived the project. H.Y., H.T., J.Y., Y.K., M.Y. performed the electric-transport, thermal-transport, and heat capacity measurements. H.Y., Y.K., M.K., K.I., T.S. performed the micro-Hall array measurements. B.R.O., Y.M.O., and S.D.W. synthesized the single crystals. All authors discussed the results and were involved in writing the paper.

Competing Interests:

The authors declare they have no competing interest.

Data and Materials availability:

All data needed to evaluate the conclusions in the paper are present in the paper and/or the Supplementary Materials.

Figures:

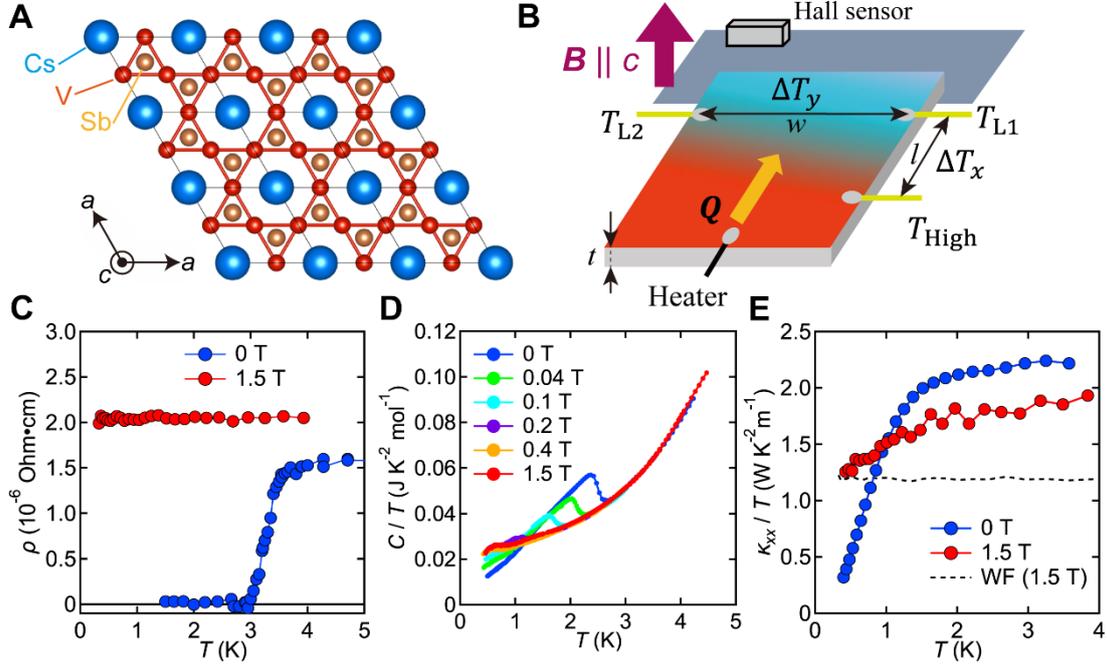

**Fig. 1 The basic properties of the kagome superconductor $CsV_3Sb_5$ and the experimental setup for the anomalous thermal Hall measurements.** (**A**) The crystal lattice structure viewed along the $c$ axis drawn by VESTA(*49*). The two-dimensional kagome structure is formed by the V–V bonds highlighted by red lines. (**B**) An illustration of the experimental setup. One heater and three thermometers ($T_{High}$, $T_{L1}$ and $T_{L2}$) were attached to the sample so that both the longitudinal ($\Delta T_x = T_{High} - T_{L1}$) and the transverse ($\Delta T_y = T_{L1} - T_{L2}$) temperature difference can be measured simultaneously as a function of the thermal current $Q$. The sample temperature $T$ is determined by $T = (T_{High} + T_{L1})/2$. Since the magnetic field applied along the $c$ axis was provided by a superconducting magnet, the zero field required for the anomalous thermal Hall measurements was achieved by monitoring the Hall sensor placed near the sample. (**C-E**) The temperature dependence of the resistivity (**C**), the heat capacity divided by the temperature (**D**) and the longitudinal thermal conductivity divided by the temperature (**E**). The dashed line in (**E**) shows the electronic contribution in $\kappa_{xx}/T$ given by the resistivity in the normal state (1.5 T in (**C**)) through the Wiedemann-Franz law.

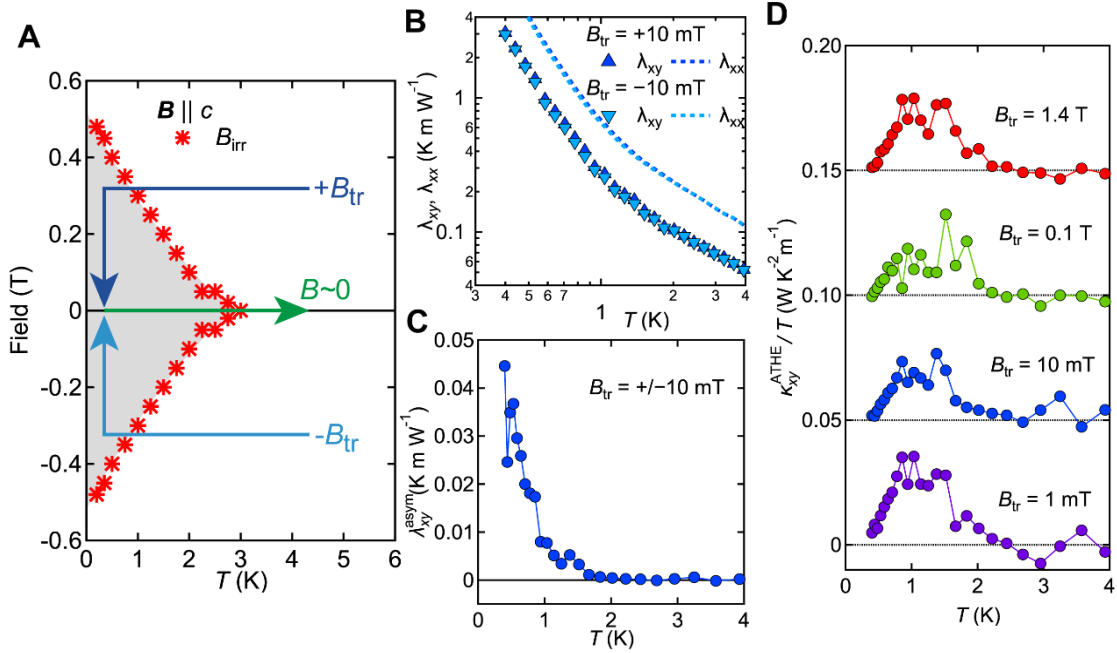

**Fig. 2 Anomalous thermal Hall measurements**. (**A**) The measurement sequence used in our measurements. The grey region shows the superconducting phase enclosed by the irreversibility field ($B_{\text{irr}}$, asterisks) determined by the onset of the zero resistance in the field or the temperature dependence of the resistivity (see section S1 in SM for more details). After cooling the sample down to the lowest temperature at a finite training field ($+B_{\text{tr}}$), the magnetic field was set to zero (the blue arrow), which was followed by the thermal-transport measurements with increasing the temperature at the zero field (the green arrow). This procedure was repeated with the opposite training field ($-B_{\text{tr}}$) to antisymmetrize the data (the cyan arrow). (**B**) The temperature dependence of the thermal Hall resistivity ($\lambda_{xy}$, triangles) and the longitudinal thermal resistivity ($\lambda_{xx}$, dashed lines) measured after cooling at the training field of $\pm 10$ mT. (**C**) The temperature dependence of the anti-symmetrized thermal Hall resistivity ($\lambda_{xy}^{\text{asym}}$) obtained from the data in (**B**). (**D**) The temperature dependence of the anomalous thermal Hall conductivity divided by the temperature measured after the cooling at $B_{\text{tr}} = 1$ mT, 10 mT, 0.1 T, and 1.4 T. The data is vertically shifted for clarity.

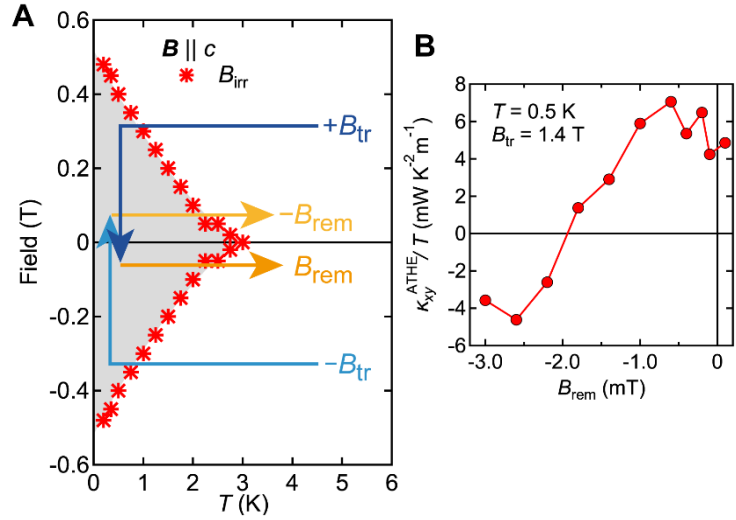

**Fig. 3 The remnant field dependence of the anomalous thermal Hall conductivity.**
(**A**) The measurement sequence used in the remnant field ($B_{\text{rem}}$) dependence measurements. After cooling the sample down to the lowest temperature at a finite training field ($+B_{\text{tr}}$), the magnetic field was set to $B_{\text{rem}}$, which was followed by the thermal-transport measurements with increasing the temperature at $B_{\text{rem}}$ (the orange arrow). This procedure was repeated with the opposite training field ($-B_{\text{tr}}$) and at the opposite remnant field ($-B_{\text{rem}}$) to antisymmetrize the thermal Hall resistance ($\lambda_{xy}(T, +B_{\text{tr}}, +B_{\text{rem}})$) as $\lambda_{xy}^{\text{asym}} = [\lambda_{xy}(T, +B_{\text{tr}}, +B_{\text{rem}}) - \lambda_{xy}(T, -B_{\text{tr}}, -B_{\text{rem}})]/2$. (**B**) The remnant field dependence of $\kappa_{xy}^{\text{ATHE}}/T$ at 0.5 K observed after cooling at $B_{\text{tr}} = 1.4$ T.

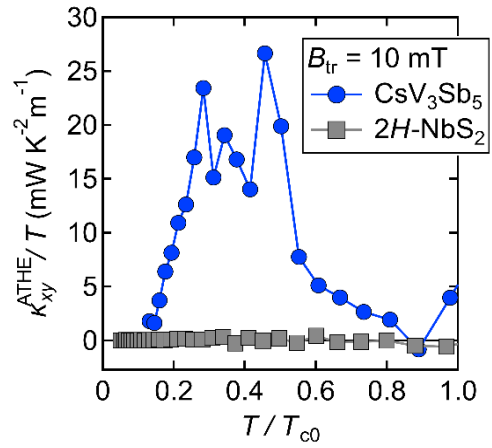

**Fig. 4 A comparison of the temperature dependence of $\kappa_{xy}^{\mathrm{ATHE}}/T$ of CsV$_3$Sb$_5$ and 2$H$–NbS$_2$.** Both data were obtained after the cooling at $B_{\mathrm{tr}} = 10$ mT. The horizontal axis is normalized by the superconducting transition temperature of each material (see section S4 for more experimental data of 2$H$–NbS$_2$).

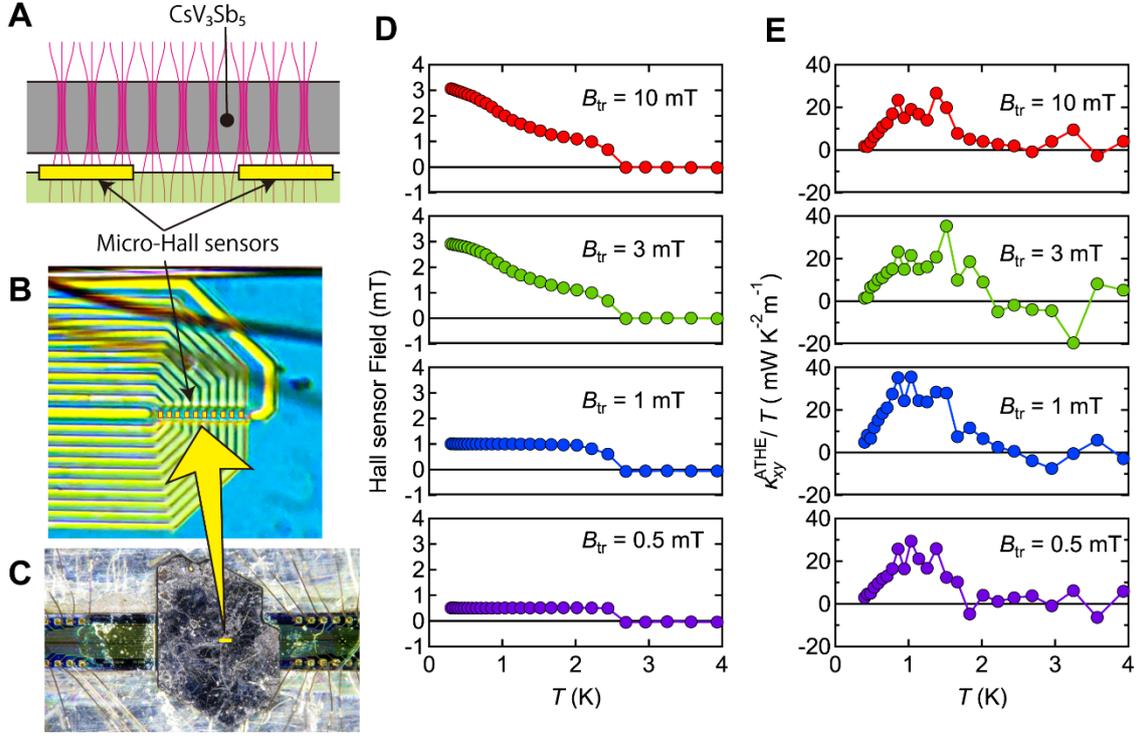

**Fig. 5 Micro-Hall array measurements.** (**A**) An illustration of the setup. The trapped magnetic field by pinned fluxes (pink lines) in the sample at zero field was detected by the micro-Hall sensors placed just below the sample. (**B**,**C**) Pictures of the micro-Hall array (**B**) and the sample set on top of the sensors (**C**). The position of the sensors is marked by the yellow rectangles. (**D**) The temperature dependence of the trapped magnetic field measured at the micro-Hall sensor located almost the center of the sample. This measurement was performed at zero field after cooling at $B_{tr}$ =0.5, 1, 3, and 10 mT, as done for the anomalous thermal Hall measurements. (**E**) The temperature dependence of the anomalous thermal Hall conductivity recorded at the same $B_{tr}$ in the micro-Hall array measurements. The data at 1 mT and 10 mT is the same with that shown in Fig. 2**D**.

Table of contents for Supplementary Materials

Section S1–S7
Figs. S1 to S7
Table S1

Section S1: Additional resistivity data of $CsV_3Sb_5$ sample #1

Figure S1a shows the temperature dependence of the resistivity near the superconducting transition. The transition temperature at onset ($T_{c,onset}$), at the middle ($T_{c,mid}$) and at the zero resistivity ($T_{c0}$) are determined as indicated by the grey symbols in Fig. S1a. The temperature dependence of the irreversibility field (Fig. 2a in the main text) is determined by the onset of the zero resistance in the field or the temperature dependence of the resistivity at different temperatures shown in Fig. S1b.

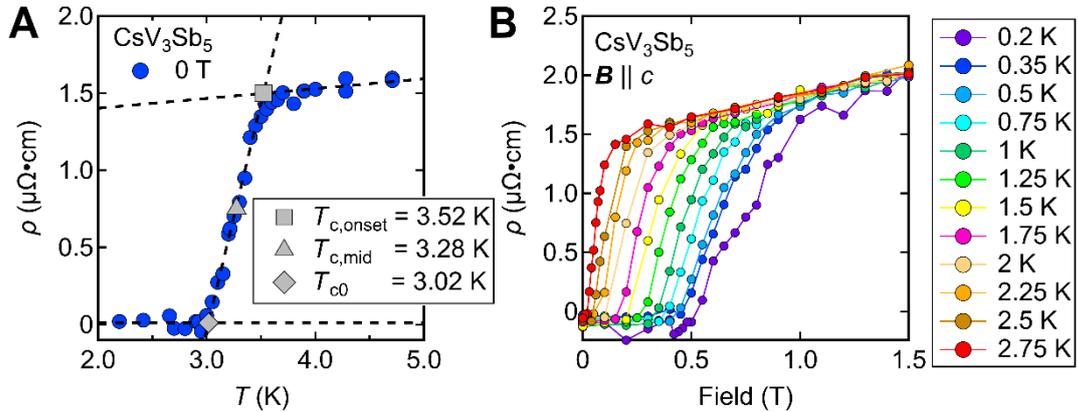

**Fig. S1 Additional resistivity data of $CsV_3Sb_5$.** (**A**) The temperature dependence of the resistivity near the superconducting transition. (**B**) The field dependence of the resistivity at different temperatures. The temperature dependence of the irreversibility field (Fig. 2**A** in the main text) is determined by the field dependence of $T_{c0}$ at each temperature.

Section S2: Thermal modeling for the mixing between the longitudinal and the transverse temperature differences.

Here, we describe a thermal modeling for the mixing between the longitudinal and the transverse temperature differences. As shown in Fig. 2B in the main text, the temperature dependence of the thermal Hall resistivity $\lambda_{xy} = t \cdot \Delta T_y/Q$ is dominated by the mixing of the longitudinal thermal resistivity $\lambda_{xx} = (wt/\ell) \cdot \Delta T_x/Q$ about 40%. This is due to a misplacement effect of the thermal contacts and the irregular shape of the sample. Suppose that we have a perfect rectangular sample with a thermal gradient applied along the $x$ axis in which the thermal contacts connected to the thermometers are slightly displaced as shown in Fig. S2. In this case, the longitudinal and the transverse temperature difference are observed as

$$\Delta T_x \equiv T_{\text{High}} - T_{L1} = \delta T_x - \frac{\beta}{w}\delta T_y \tag{S1}$$

$$\Delta T_y \equiv T_{L1} - T_{L2} = \delta T_y - \frac{\alpha}{\ell}\delta T_x, \tag{S2}$$

respectively, where $\delta T_x$ ($\delta T_y$) is the field-symmetric (field-antisymmetric) temperature difference along the $x$ ($y$) axis. Therefore,

$$\lambda_{xy} = t \cdot \frac{\Delta T_y}{Q} = t \cdot \frac{\delta T_y}{Q} - \frac{\alpha}{w}\left(\frac{w \cdot t}{\ell}\right)\frac{\delta T_x}{Q} \tag{S3}$$

contains a mixing component from $\delta T_x$ and so does $\lambda_{xx}$ from $\delta T_y$ as

$$\lambda_{xx} = \left(\frac{w \cdot t}{\ell}\right)\frac{\Delta T_x}{Q} = \left(\frac{w \cdot t}{\ell}\right)\frac{\delta T_x}{Q} - \frac{\beta}{\ell} \cdot t\frac{\delta T_y}{Q}. \tag{S4}$$

In reality, the shape of the sample is very different from a perfect rectangle (see Fig. 5C in the main text), tilting the direction of the thermal gradient from the direction parallel to the sample edge whereas the two thermal contacts for reading $T_{L1}$ and $T_{L2}$ are placed assuming that the thermal gradient is perpendicular to the edge of the sample. Given the sample width of about 1.8 mm and the finite size of the thermal contacts (~ 0.2 mm), a mixing of 40% (displacement of $\alpha \sim \pm 0.4$ mm) will often be unavoidable.

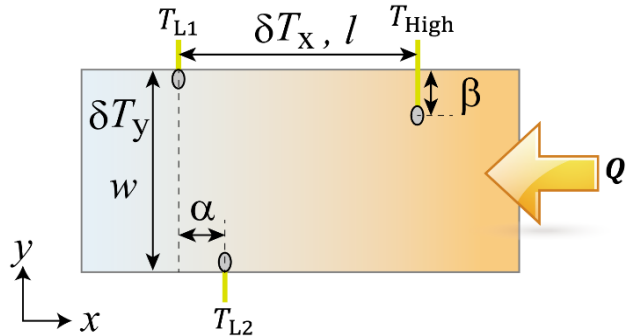

**Fig. S2 An illustration of a thermal modeling of the thermal-transport**

**measurements.** The mixings of the longitudinal ($\delta T_x$) and transverse ($\delta T_y$) temperature differences into the observed ones ($\Delta T_x = T_{\text{High}} - T_{\text{L1}}$ and $\Delta T_y = T_{\text{L1}} - T_{\text{L2}}$) are caused by the displacements of the thermal contacts ($\alpha$ and $\beta$) from the perfect diagonal location.

We note that the field-antisymmetric component is observed not only in $\lambda_{xy}$ but also in $\lambda_{xx}$. As shown in Fig. S3A, the anti-symmetrized component in $\lambda_{xx}$ ($\lambda_{xx}^{\text{asym}}$, open symbols) shows a similar temperature dependence with $\lambda_{xy}^{\text{asym}}$ with a larger magnitude. To explain $\lambda_{xx}^{\text{asym}}$ by a mixing of $\lambda_{xy}^{\text{asym}}$ due to the misalignment effect, we need to assume $\beta > \ell$, which is unlikely even considering an inclination of the direction of the heat current from the $x$ axis. This large $\lambda_{xx}^{\text{asym}}$ exceeding $\lambda_{xy}^{\text{asym}}$ could be attributed to the presence of an antisymmetric $\delta T_x$, which in turn might be mixed into $\lambda_{xy}^{\text{asym}}$.

To clarify this issue, we first excluded the possibility of an experimental artifact that the antisymmetric components come from the thermometers used in this study, by measuring the thermal conductivity of a gold wire with 50 μm diameter as shown in Fig. S3D. The temperature dependence of $\Delta T_x/Q$ and $\Delta T_y/Q$ of the gold wire was checked by anti-symmetrizing them with respect to the training field as done for the thermal Hall measurements of CsV$_3$Sb$_5$. As shown in Fig. S3E, both the anti-symmetrized $\Delta T_x/Q$ and $\Delta T_y/Q$ are confirmed to be much smaller than those observed in CsV$_3$Sb$_5$ (Fig. S3F), showing that neither $\lambda_{xx}^{\text{asym}}$ nor $\lambda_{xy}^{\text{asym}}$ are due to an experimental artifact.

Next, we repeated the measurements of CsV$_3$Sb$_5$ by varying the position of thermometers to reduce the mixing between them. In this revised setup, the mixing of $\lambda_{xy}/\lambda_{xx}$ (i.e. $\alpha/w$ in the equation of $\lambda_{xy}$ above) was reduced to about 20% from about 40% of the initial measurements (Fig. 2B in the main text). Despite this change in the setup, we find that $\lambda_{xy}^{\text{asym}}$ is well reproduced as shown in Fig. S3B. This good reproducibility of $\lambda_{xy}^{\text{asym}}$ indicates that $\lambda_{xy}^{\text{asym}}$ reflects the anomalous Hall effect (i.e. $\delta T_y$) and is not caused by a mixing of $\lambda_{xx}^{\text{asym}}$.

On the other hand, although $\lambda_{xx}^{\text{asym}}$ was reduced in about half in this revised setup, $\lambda_{xx}^{\text{asym}}$ is still larger than $\lambda_{xy}^{\text{asym}}$. Furthermore, the mixing of $\lambda_{xy}$ into $\lambda_{xx}$ (i.e. $\beta/\ell$ in the equation above) was checked in this revised setup by measuring the normal thermal Hall signal in the normal state above 2 T at 1 K. As shown in Fig. S3C, the mixing ratio is found as $-14\%$, which is not only too small to explain $\lambda_{xx}^{\text{asym}}$ from $\lambda_{xy}^{\text{asym}}$ but also has the opposite sign. These results indicate that $\lambda_{xx}^{\text{asym}}$ contains an antisymmetric component that cannot be explained solely by a misalignment effect of the transverse component. Clarifying the mechanism behind $\lambda_{xx}^{\text{asym}}$ remains a challenge for the future.

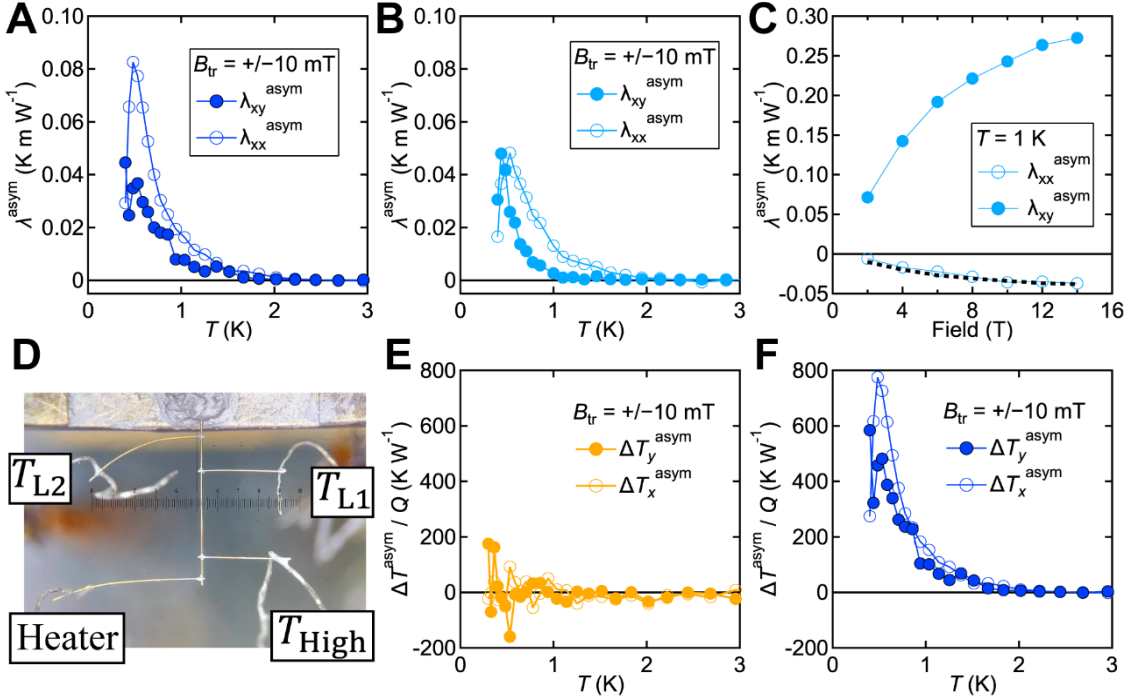

Fig. S3 (A,B) The temperature dependence of $\lambda_{xy}^{\text{asym}}$ (filled circles) and $\lambda_{xx}^{\text{asym}}$ (open circles) anti-symmetrized with respect to the training field of 10 mT of CsV$_3$Sb$_5$ (sample #1) recorded in the first run (A) and the additional run after adjusting the position of the thermal contracts (B). (C) The field dependence of $\lambda_{xy}^{\text{asym}}$ and $\lambda_{xx}^{\text{asym}}$ in the normal state at 1 K. The black dashed line shows the data of $\lambda_{xy}^{\text{asym}}$ multiplied by $-0.14$. (D) a picture showing the setup of the gold wire measurement. All the thermometers and the heater were connected to the gold wire with a diameter of 50 μm by a silver paste. (E) The temperature dependence of $\Delta T_x = T_{\text{High}} - T_{\text{L1}}$ (open circles) and $\Delta T_y = T_{\text{L1}} - T_{\text{L2}}$ (filled circles) of the gold wire. The data is anti-symmetrized with respect to the training field of $B_{\text{tr}} = \pm 10$ mT. (F) The same data of (A) shown in the unit of $\Delta T / Q$.

Section S3: Reproducibility of the anomalous thermal Hall effect

The reproducibility of the anomalous thermal Hall effect was checked by performing the same measurements in sample #2. As shown in Fig. S3, sample #2 shows the lower resistivity (Figs. S4A and S4B) and the higher thermal conductivity (Fig. S4C). Although the scatter of the anomalous thermal Hall conductivity is higher due to this higher thermal conductivity, a good reproducibility is confirmed in both the anomalous thermal Hall conductivity (Figs. S4D–G) and the thermal Hall angle (Figs. S4H–K).

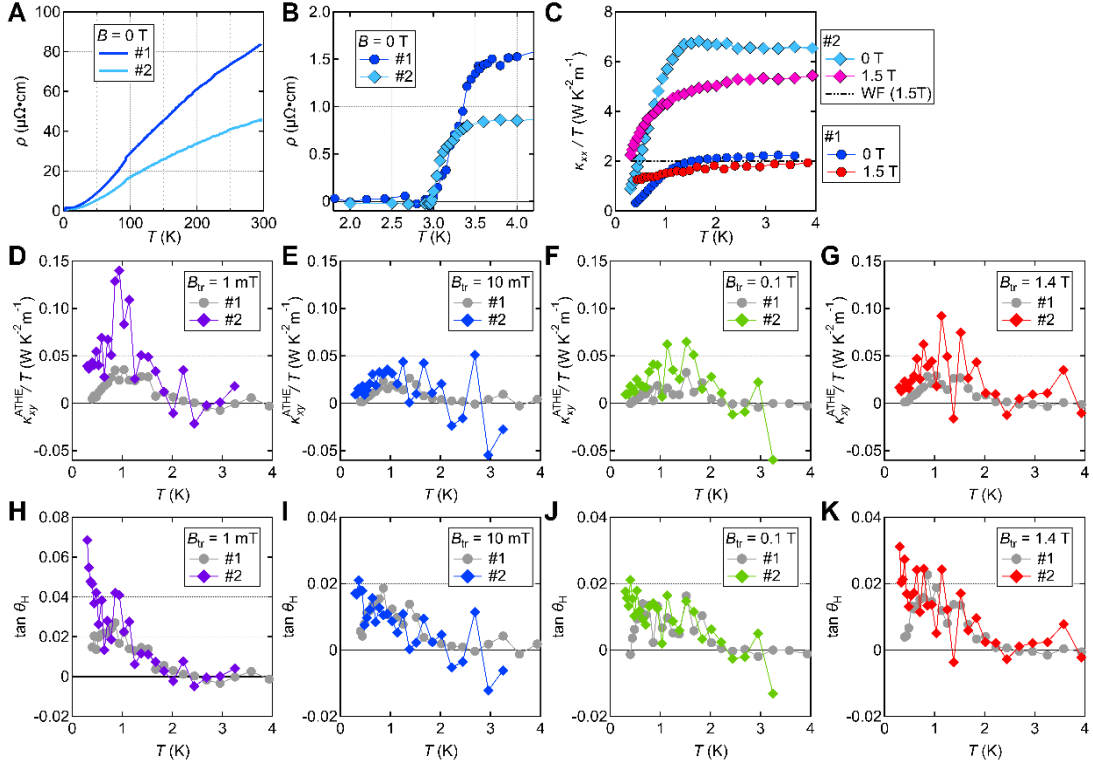

**Fig. S4 Reproducibility of the spontaneous thermal Hall effect observed in sample #2**. (**A**–**B**) The temperature dependence of the resistivity in the whole temperature range (**A**) and near the superconducting transition temperature (**B**) of sample #1 and sample #2. (**C**) The temperature dependence of the longitudinal thermal conductivity divided by the temperature of sample #2 (diamonds) at 0 T and in the normal state at 1.5 T, shown with the data of sample #1 (the same data shown in Fig. 1**E** in the main text). The value of $\kappa_{xx}/T$ estimated by the resistivity at 1.5 T through the Wiedemann-Franz law is shown by the dashed line. (**D**–**K**) The temperature dependence of the anomalous thermal Hall conductivity (**D**–**G**) and that of the thermal Hall angle $\tan\theta_\mathrm{H} = \kappa_{xy}/\kappa_{xx}$ (**H**–**K**) obtained after cooling with the training field at 1 mT, 10 mT, 0.1 T and 1.4 T. The data of sample #1 (the same data in Fig. 3 in the main text) is shown in grey for reference.

Section S4: Anomalous thermal Hall measurements in 2H-NbS$_2$

Our experimental setup for the anomalous thermal Hall measurements was examined by performing the same measurements in the conventional type-II superconductor 2H-NbS$_2$ by using the same setup used in the measurements of CsV$_3$Sb$_5$. Figures S5A–C show the temperature dependence of the resistivity and the thermal conductivity of 2H-NbS$_2$. As shown in Fig. 4 in the main text and Fig. S5D, both the anomalous thermal Halll conductivity and the thermal Hall angle $\tan\theta_H = \kappa_{xy}^{\mathrm{ATHE}}/\kappa_{xx}$ show the negligibly small signal compared to those in CsV$_3$Sb$_5$, ensuring that the observed anomalous thermal Hall effect in CsV$_3$Sb$_5$ is not an artifact but an inherent property of the superconducting state of CsV$_3$Sb$_5$.

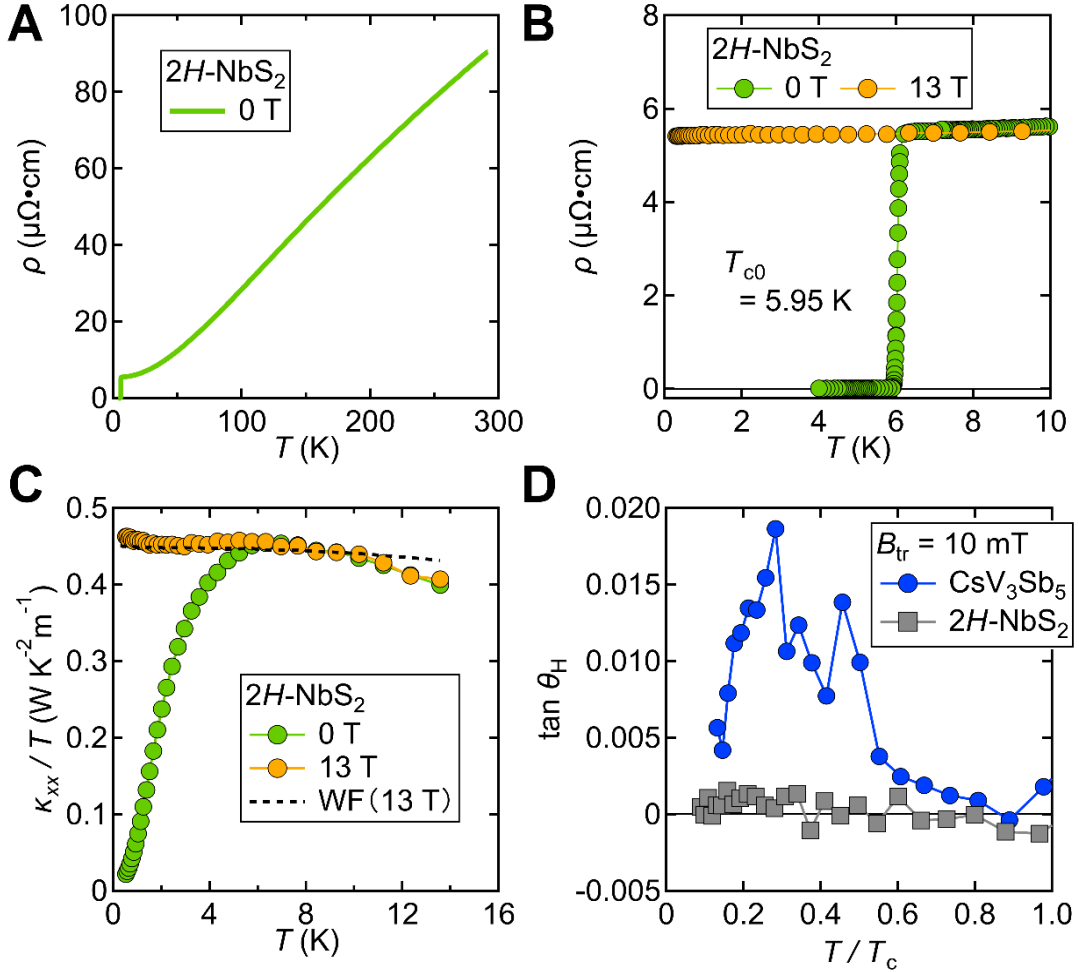

**Fig. S5 Additional data of 2H–NbS$_2$.** (**A**) The temperature dependence of the resistivity at 0 T. (**B**) The temperature dependence of the resistivity near the superconducting transition at 0 T and at 13 T. (**C**) The temperature dependence of the longitudinal thermal

conductivity divided by the temperature at 0 T and at 13 T. The value of $\kappa_{xx}/T$ estimated by the resistivity at 13 T through the Wiedemann-Franz law is shown by the dashed line. **(D)** The temperature dependence of the thermal Hall angle $\tan\theta_H = \kappa_{xy}^{\text{ATHE}}/\kappa_{xx}$ of CsV3Sb5 #1 (blue circles) and 2$H$–NbS2 (grey rectangles) obtained after cooling the training field of 10 mT.

Section S5: Additional micro-Hall array measurements with thermal gradient

To investigate a thermal response of the trapped vortices in the superconducting state of CsV3Sb5, additional micro-Hall array measurements were performed by attaching a heater, two thermometers ($T_{\text{High}}$ and $T_{\text{L}}$) and a contact to the ground (Fig. S6A) so that a temperature gradient ($\Delta T_x = T_{\text{High}} - T_{\text{L}}$) could be applied to the sample during the trapped field measurements. These measurements were performed at zero field after cooling at the training field of $B_{\text{tr}} = \pm 0.5, \pm 1, \pm 3,$ and $\pm 10$ mT (up and down triangles in Figs. S6B–E for the positive and negative $B_{\text{tr}}$, respectively) with a thermal gradient of $\Delta T_x/T = 2.5\%$ (filled symbols) and without the thermal current (open symbols), where the sample temperature $T = (T_{\text{High}} + T_{\text{L}})/2$. As shown in Figs. S6B–E, the trapped field observed with and without the thermal gradient is virtually the same, confirming that the trapped vortices are not de-pinned by the thermal gradient applied during the thermal-Hall measurements. The slight decrease of the trapped field from the results shown in Fig. 5D in the main text is attributed to a change of the position of the micro-Hall sensor with respect to the sample. Interestingly, although a symmetric trapped field is confirmed at $B_{\text{tr}} = \pm 0.5$ mT, a slight asymmetric trapped field is observed at higher training fields near $T_c$, which might be related to the broken time-reversal-symmetry in the superconducting state. Nevertheless, the fact that $\kappa_{xy}^{\text{ATHE}}$ does not depend on $B_{\text{tr}}$ even though the trapped field clearly depends on $B_{\text{tr}}$ shows that $\kappa_{xy}^{\text{ATHE}}$ is not caused by the trapped vortices.

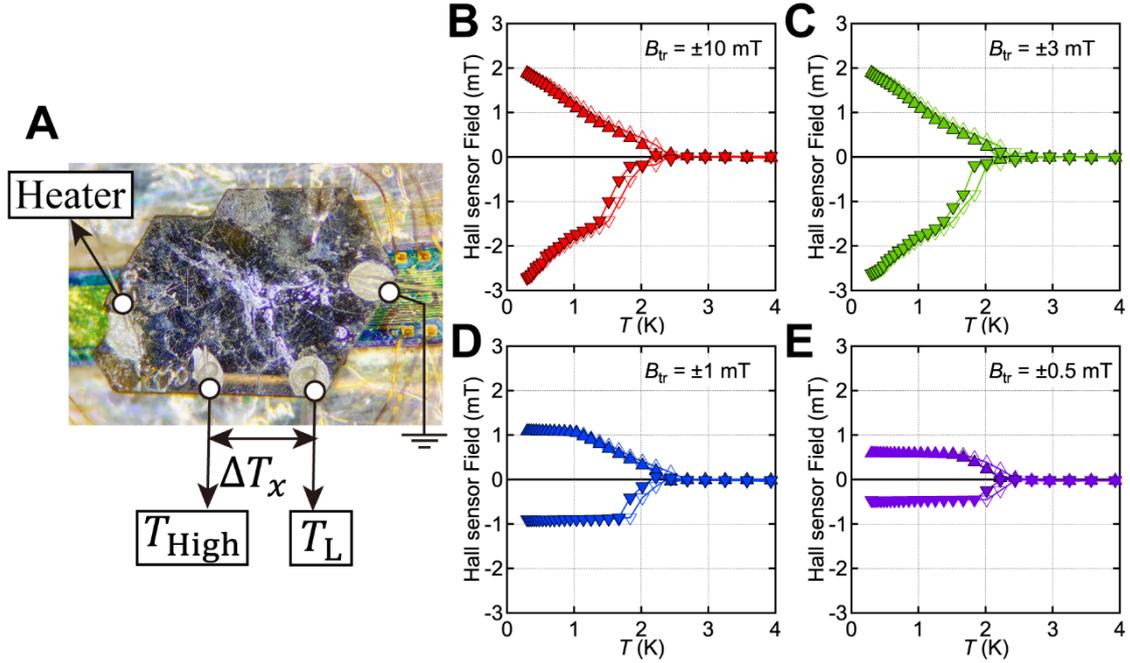

**Fig. S6 Additional micro-Hall array measurements with thermal gradient.** (**A**) a picture showing the sample placed on top of the micro-Hall sensor in the same manner shown in Fig. 5C in the main text. (**B**–**E**) The temperature dependence of the trapped magnetic field measured by the micro-Hall sensor located almost the center of the sample at zero field after cooling at the training field of $B_{tr} = \pm 0.5, \pm 1, \pm 3,$ and $\pm 10$ mT (up and down triangles for the positive and negative $B_{tr}$, respectively) with a thermal gradient of $\Delta T_x/T = 2.5\%$ (filled symbols) and without the thermal current (open symbols).

Section S6: The temperature dependence of the anomalous thermal Hall conductivity observed at the different remnant fields

Figure S7 shows the temperature dependence of the anomalous thermal Hall conductivity measured at a finite remnant field ($B_{rem}$) after cooling the sample following the procedure shown in Fig. 3A in the main text. This result was obtained in an early run before installing the Hall sensor to monitor the magnetic field around the sample (Fig. 1B in the main text). During the measurement, the training field was turned off by simply setting the output of the magnet power supply to zero, which was lately found to leave a remnant field of which the magnitude depends on the training field as the trapped field of the superconducting magnet. The magnitude of the remnant field is estimated by repeating the same field procedure after installing the Hall sensor. As shown in Fig. S7, the temperature dependence of the anomalous thermal Hall conductivity is completely reversed when $|B_{\rm rem}| > 2$ mT, implying the reversal of the polarity of the chiral domains.

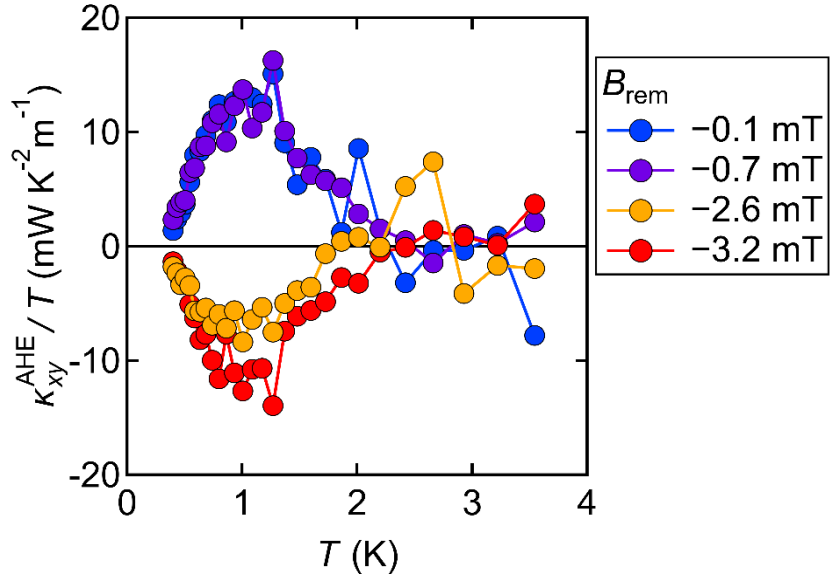

**Fig. S7 The temperature dependence of the anomalous thermal Hall conductivity observed at the different remnant fields ($B_{\rm rem}$).**

Section S7: The estimation of $\ell_{\rm mf}/\xi_0$ from the residual resistivity

The ratio of the mean free path $\ell_{\rm mf}$ to the coherence length $\xi_0$ is estimated from

$$\frac{\xi_0}{\ell_{\rm mf}} = \frac{\hbar}{\pi \tau \Delta(0)}, \tag{S5}$$

where $\tau$ is the scattering time and $\Delta(0)$ the superconducting gap at 0 K. The scattering time is given by the vacuum magnetic permeability $\mu_0$, the London penetration depth at 0 K $\lambda_L(0)$, and the residual resistivity $\rho_0$ as

$$\tau = \frac{\mu_0 \lambda_L^2(0)}{\rho_0}. \tag{S6}$$

The residual resistivity of our sample is estimated as 1.3 μΩ·cm and 0.7 μΩ·cm for the sample #1 and #2, respectively (Fig. S5B). By using the values of $\Delta(0) = 0.33$ meV and $\lambda_L(0) = 387$ nm from Ref. (*36*), $\ell_{\text{mf}}/\xi_0$ is estimated as 23 and 43 for the sample #1 and #2, respectively. These estimations are in the same order of $\ell_{\text{mf}}/\xi_0 = 7.5$ used in the calculation of $\kappa_{xy}^{\text{ATHE}}$ in Ref. (*14*).

Table S1 Sample dimensions used in this study.

|  | Length (mm) | Width (mm) | Thickness (mm) |
| --- | --- | --- | --- |
| CsV$_3$Sb$_5$ #1 | 1.34 | 1.87 | 0.076 |
| CsV$_3$Sb$_5$ #2 | 2.15 | 1.78 | 0.077 |
| 2$H$-NbS$_2$ | 1.62 | 1.41 | 0.041 |